\documentclass[twocolumn,showpacs,preprintnumbers,aps,prb]{revtex4-1}

\usepackage{graphicx,color}

\usepackage{dcolumn}

\usepackage{bm}

\begin{document}

\preprint{}

\title{Partially disordered state and spin-lattice coupling in an $S$=3/2 triangular lattice antiferromagnet Ag$_2$CrO$_2$}

\author{M. Matsuda}

\affiliation{Quantum Condensed Matter Division, Oak Ridge National Laboratory, Oak Ridge, Tennessee 37831, USA}

\author{H. Yoshida}

\affiliation{National Institute for Materials Science (NIMS), 1-1 Namiki, Tsukuba, Ibaraki 305-0044, 
Japan}

\author{M. Isobe}

\affiliation{National Institute for Materials Science (NIMS), 1-1 Namiki, Tsukuba, Ibaraki 305-0044, 
Japan}

\author{C. de la Cruz}

\affiliation{Quantum Condensed Matter Division, Oak Ridge National Laboratory, Oak Ridge, Tennessee 37831, USA}

\author{R. S. Fishman}

\affiliation{Materials Science and Technology Division, Oak Ridge National Laboratory, Oak Ridge, Tennessee 37831, USA}

\date{\today}

\begin{abstract}

Ag$_2$CrO$_2$ is an $S$=3/2 frustrated triangular lattice antiferromagnet without orbital degree of freedom. With decreasing temperature, a 4-sublatice spin state develops. However, a long-range partially disordered state with 5 sublattices abruptly appears at $T\rm_N$=24 K, accompanied by a structural distortion, and persists at least down to 2 K. The spin-lattice coupling stabilizes the anomalous state, which is expected to appear only in limited ranges of further-neighbor interactions and temperature. It was found that the spin-lattice coupling is a common feature in triangular lattice antiferromagnets with multiple-sublattice spin states, since the triangular lattice is elastic.

\end{abstract}


\maketitle

\section{Introduction}
Geometrically frustrated magnets give rise to many interesting phenomena originating from the macroscopic ground state degeneracy.~\cite{Diep} Spin liquid ground state is expected in strongly frustrated quantum kagom\'{e} and pyrochlore lattice Heisenberg antiferromagnets.~\cite{moes98,canals00,yan11} 
Although quantum triangular lattice Heisenberg antiferromagnet shows 120$^{\circ}$ long-range magnetic order with 3 sublattices,~\cite{bermu94} additional multiple-spin interactions can lead to spin liquid state.~\cite{liming00} Triangular lattice Heisenberg antiferromagnets with classical spins also show interesting phenomena related to the chirality degree of freedom. A magnetic transition driven by the binding-unbinding of the $Z_2$ vortices is predicted to occur at a temperature, where the vortex correlation length diverges but the spin correlation length remains finite.~\cite{kawa84,kawa10}

Spin-lattice coupling sometimes plays an important role in selecting the ground state in the frustrated magnets. Even if there is no orbital degree of freedom, a small amount of structural distortion is sufficient to lift the ground state degeneracy and stabilize a long-range magnetic order.
For example, in the Cr-based spinels, consisting of the pyrochlore lattice, the magnetic order is observed far below the Curie-Weiss temperature $\Theta_{CW}$ due to spin-lattice coupling.~\cite{lee00}
Furthermore, a half-magnetization plateau is stabilized in a wide range of magnetic field by the spin-lattice coupling.~\cite{ueda05,penc04} Similar effect is also predicted for the triangular and kagom\'{e} lattice antiferromagnets.~\cite{wang08}

Ag$_2M$O$_2$ ($M$=Cr, Mn, and Ni) consists of triangular lattice planes of $M$O$_2$, which are well separated by the metallic Ag$_2$ layers.~\cite{yoshida06,yoshida08} Ag$_2$NiO$_2$ (Ni$^{3+}$, $d^7$ low spin $S$=1/2) and Ag$_2$MnO$_2$ (Mn$^{3+}$, $d^4$ high spin $S$=2), which have orbital degree of freedom, were studied previously.~\cite{yoshida06,yoshida08,nozaki08,ji10} Both compounds show Jahn-Teller distortions, which affect the magnetic ground state. In Ag$_2$NiO$_2$, a long-range magnetic order with $Q_m$=($\frac{1}{3}$, $\frac{1}{3}$, 0)$\rm_{hexagonal}$ occurs below N\'{e}el temperature $T\rm_N$=56 K. In Ag$_2$MnO$_2$, a glassy state with a short-range magnetic order with $Q_m$=($\frac{1}{2}$, $\frac{1}{2}$, 0)$\rm_{monoclinic}$ is observed as the ground state.

\begin{figure}
\includegraphics[width=8.2cm]{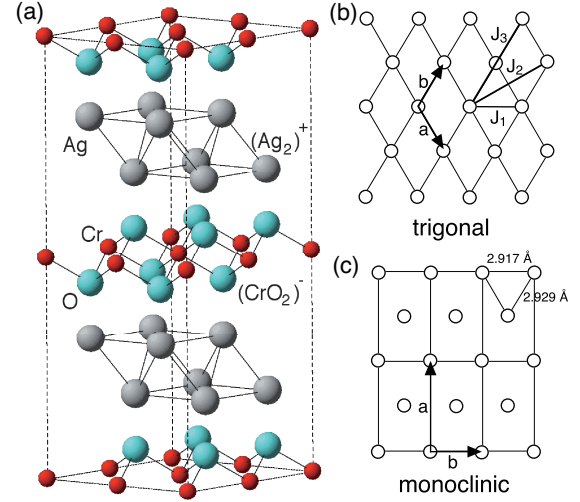}
\caption{(Color online) (a) Crystal structure of Ag$_2$CrO$_2$ in trigonal ($P\bar{3}m1$) phase above $T\rm_N$ (=24 K). 2$\times$2$\times$2 unit cells are shown. Note that there is only one CrO$_2$ triangular lattice plane in a single unit cell. (b) and (c) show schematic structures of the triangular lattice plane of the Cr$^{3+}$ spins in trigonal ($T>T\rm_N$) and monoclinic phases ($T<T\rm_N$), respectively. Magnetic interactions ($J_1$, $J_2$, and $J_3$) are shown in (b).}
\label{structure}
\end{figure}
Ag$_2$CrO$_2$ has a trigonal structure ($P\bar{3}m1$) with $a$=2.9298 \AA\ and $c$=8.6637 \AA\ at $T$=200 K, as shown in Fig. \ref{structure}(a), and consists of regular triangular lattices of the magnetic Cr$^{3+}$ (nominally $t_{2g}^3; S = 3/2)$ ions without orbital degree of freedom. Therefore, Ag$_2$CrO$_2$ is a good candidate for a model compound of a regular triangular lattice Heisenberg antiferromagnet. The specific heat shows a sharp peak at $T$=24 K.~\cite{yoshida11} The magnetic susceptibility shows that $\Theta_{CW}$ = $-$97 K and the effective moment $p\rm_{eff}$ = 3.55 $\mu\rm_B$, consistent with 3.87 $\mu\rm_B$ expected for $S$=3/2.~\cite{yoshida11}  These results indicate that an antiferromagnetic long-range order develops below $T\rm_N$=24 K. In addition, the magnetic susceptibility shows weak ferromagnetic component below $T\rm_N$.

We performed neutron diffraction experiments on a powder sample of Ag$_2$CrO$_2$. It was found that Ag$_2$CrO$_2$ shows a partially disordered (PD) state with 5 sublattices and a structural distortion simultaneously below $T\rm_N$=24 K, indicating a spin-lattice coupling to stabilize the N\'{e}el order. We also measured the magnetic diffuse scattering and found that the material does not prefer the 5-sublattice (5SL) structure but 4-sublattice (4SL) structure above $T\rm_N$. The structural distortion is considered to modify further-neighbor interactions so that the 5SL structure becomes stable.
As far as we know, Ag$_2$CrO$_2$ is the first quasi-two-dimensional (2D) triangular lattice antiferromagnet that shows the long-range PD state.

\section{Experimental Details}
A powder sample of Ag$_2$CrO$_2$ was prepared by the solid state reaction technique with stoichiometric mixture of Ag, Ag$_2$O, and Cr$_2$O$_3$ powder in high pressure.~\cite{yoshida11} The powder sample that weighs $\sim$2.5 g was used. The neutron powder diffraction experiments were carried out on a neutron powder diffractometer HB-2A, installed at HFIR in Oak Ridge National Laboratory (ORNL). We utilized two wavelengths ($\lambda$) 1.5374 and 2.410 \AA\ for structural analysis and high-resolution measurement, respectively. Magnetic diffuse scattering was measured on a thermal triple-axis neutron spectrometer HB-1, installed at HFIR in ORNL. The horizontal collimator sequence was 48'-80'-S-80'-120'. The fixed incident neutron energy was 13.5 meV with the energy resolution ($\Delta E$) of 1.2 meV. Contamination from higher-order beams was effectively eliminated using PG filters.

\section{Experimental Results}
\begin{figure}
\includegraphics[width=8.0cm]{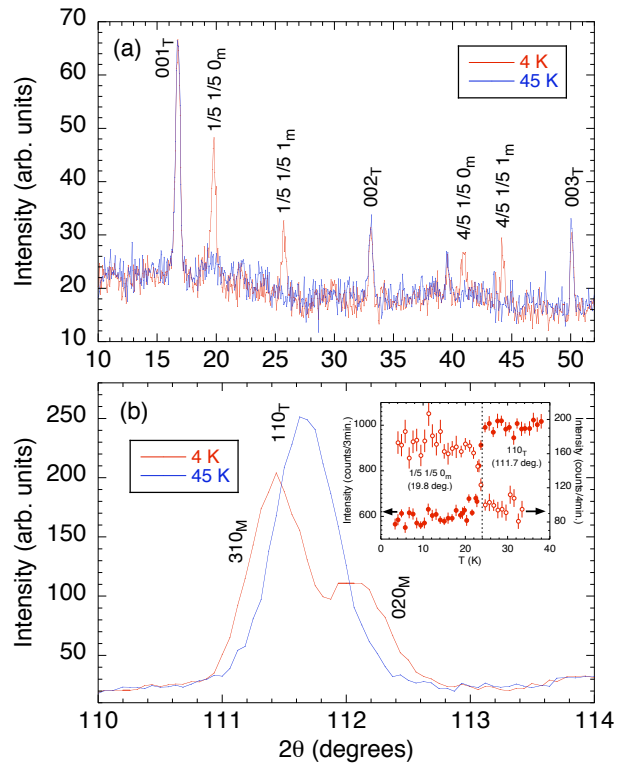}
\caption{(Color online) Neutron powder diffraction patterns in Ag$_2$CrO$_2$ at 4 and 45 K measured with $\lambda$=2.410 \AA. The data at low scattering angles show that some magnetic Bragg peaks develop below $T\rm_{N}$ (a) and those at high scattering angles show that a nuclear Bragg peak splits below $T\rm_{N}$ (b). The inset shows the temperature dependence of the magnetic intensity at 19.8$^\circ$ and nuclear intensity at 117.9$^\circ$, measured with increasing temperature. No hysteresis was observed. ``T" and ``M" denote that the indices are for nuclear reflections in trigonal and monoclinic phases, respectively. ``m" denotes that the indices are for magnetic reflections. The magnetic reflections are indexed on the basis of the trigonal structure.}
\label{profile}
\end{figure}
Figure \ref{profile} shows the neutron powder diffraction patterns in Ag$_2$CrO$_2$ at 4 and 45 K, measured with a high resolution mode ($\lambda$=2.410 \AA). The magnetic Bragg reflections, observed below $T\rm_N$, are shown in Fig. \ref{profile}(a). The magnetic reflections with $\frac{1}{5}$$\frac{1}{5}$$L$ and $\frac{4}{5}$$\frac{1}{5}$$L$, where $L$=0 and 1, were observed, indicating that the magnetic structure has a 5SL state in the triangular plane and also the unit cell along the $c$ axis is the same as the chemical one so that the magnetic arrangement is ferromagnetic along the $c$ axis. Figure \ref{profile}(b) shows that a nuclear Bragg reflection 110 splits into two below $T\rm_N$. The inset in Fig. \ref{profile}(b) shows the temperature dependence of $\frac{1}{5}$$\frac{1}{5}$0 magnetic and 110 nuclear Bragg peak intensities. The magnetic intensity abruptly develops below $T\rm_N$, whereas the nuclear intensity drops, indicating that the magnetic and structural phase transitions occurs simultaneously at $T\rm_N$. This result indicates that the spin-lattice coupling releases the highly frustrated state and gives rise to a N\'{e}el state.

\begin{figure}
\includegraphics[width=8.5cm]{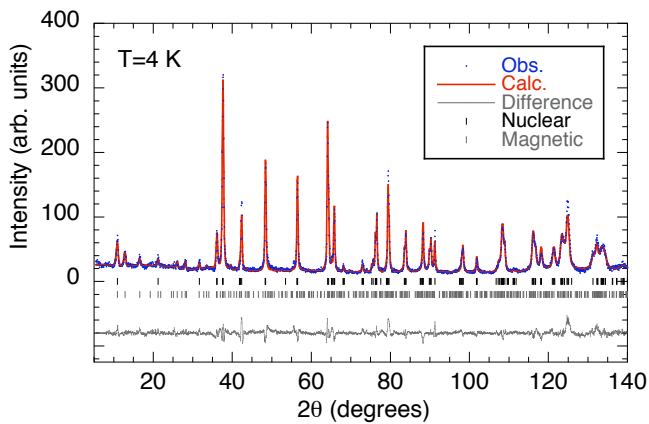}
\caption{(Color online) Rietveld refinement of the neutron powder diffraction pattern measured with $\lambda$=1.5374 \AA\ at $T$= 4 K, where nuclear and magnetic Bragg peaks coexist.}
\label{rietveld}
\end{figure}
\begin{figure}
\includegraphics[width=8.0cm]{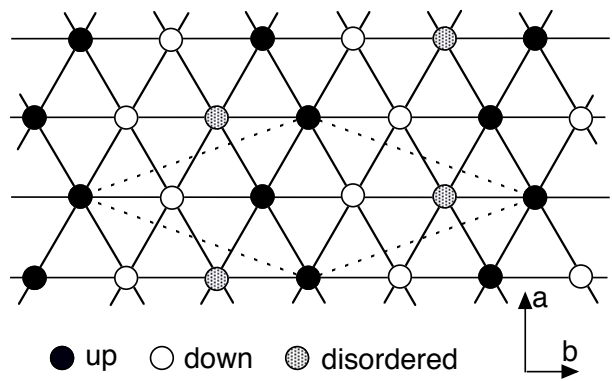}
\caption{Schematic spin structure of the PD state with 5 sublattices below $T\rm_N$. The ordered spins point along the $c$ axis almost perpendicular to the triangular plane. The dashed lines represent the magnetic unit cell.}
\label{structure2}

\end{figure}

Figure \ref{rietveld} shows the neutron powder diffraction pattern at $T$=4 K in Ag$_2$CrO$_2$, measured with $\lambda$=1.5374 \AA. We performed a Rietveld refinement, using the Fullprof package.~\cite{fullprof} In the refinement, monoclinic structure with $C2/m$ symmetry, which is a subgroup of $P\bar{3}m1$, is assumed. The schematic structures for the high-temperature trigonal and low-temperature monoclinic phases are shown in Fig. \ref{structure}(b) and \ref{structure}(c), respectively. In the monoclinic phase, the triangular lattice is slightly contracted along the $b$ axis. The structural parameters determined by the refinement are shown in Table I.
For the magnetic structure, a PD magnetic structure with 5 sublattices, shown in Fig. \ref{structure2}, is assumed. In this model spin arrangement is
up-down-up-down-disordered-..... along the $b$ axis. The spins are assumed to point along the $c$ axis with a ferromagnetic spin arrangement along this axis.
The weak ferromagnetic component observed in the magnetization measurement is consistent with the PD model since it probably originates from the small ferromagnetic moments induced at the disordered sites in magnetic field.
Both the nuclear and magnetic Bragg reflections are fitted with the models reasonably well. The Bragg $R$-factors for crystal and magnetic structures are $R\rm_{nuc}$=6.66\% and $R\rm_{mag}$=11.5\%, respectively. The ordered magnetic moment at the ordered spin sites was fitted to be 2.9(1) $\mu\rm_B$, which almost corresponds to the full moment for $S$=3/2.

\begin{table}
\caption{Structural parameters of Ag$_2$CrO$_2$ determined by the Rietveld refinements at $T$=4 and 45 K.}
\label{t1}
\raggedright
$T$=45 K, Trigonal ($P$$\bar{3}$$m$1)\\
\begin{tabular}{cccccc}
\hline
& position & $x$ & $y$ & $z$ & $U$(\AA$^2$)\\
\hline
Ag & 2$d$ & 0.3333 & 0.6667 & 0.6284(5) & 0.06(9)\\
Cr & 1$a$ & 0.0000 & 0.0000 & 0.0000 & 0.47(12)\\
O & 2$d$ & 0.6667 & 0.3333 & 0.1197(5) & 0.15(9)\\
\hline
\end{tabular}\\
$a$=2.9246140(7) \AA, $b$=2.9246140(7) \AA, $c$=8.6610155(1) \AA\\
$R\rm_{nuc}$=6.80\%\\
\vspace{5 mm}
$T$=4 K, Monoclinic ($C$2/$m$)\\
\begin{tabular}{cccccc}
\hline
& position & $x$ & $y$ & $z$ & $U$(\AA$^2$)\\
\hline
Ag & 4$i$ & 0.3375(11) & 0.0000 & 0.3719(4) & 0.05(6)\\
Cr & 2$a$ & 0.0000 & 0.0000 & 0.0000 & 0.52(10)\\
O & 4$i$ & 0.3309(12) & 0.0000 & 0.1195(4) & 0.38(7)\\
\hline
\end{tabular}\\
$a$=5.07976(16) \AA, $b$=2.916985(9) \AA, $c$=8.66164(2) \AA, $\beta$=90.0723(3)$^\circ$\\
$M\rm_{Cr}$=2.9(1) $\mu\rm_B$\\
$R\rm_{nuc}$=6.66\%, $R\rm_{mag}$=11.5\%
\end{table}

The structural distortion, which is accompanied by the magnetic transition, is unexpected because the Cr$^{3+}$ ions have no orbital degree of freedom. This behavior is very similar to that observed in the highly-frustrated Cr-based spinel $A$Cr$_2$O$_4$ ($A$=Zn, Cd, and Hg), in which the spin-lattice coupling stabilizes the magnetic structure both in ambient and high magnetic field.~\cite{lee00,ueda05}

It is also reported that a triangular lattice antiferromagnet CuFeO$_2$, in which Fe$^{3+}$ ion with $S$=5/2 has no orbital degree of freedom, shows a very similar behavior.~\cite{ye06,terada06}
In CuFeO$_2$, a monoclinic lattice distortion elongates the triangular lattice along the $b$ axis and the spin structure is a 4SL state (up-down-up-down-..... along the $b$ axis). Since the antiferromagnetic coupling between nearest-neighbor (NN) Fe$^{3+}$ spins is mediated by the Fe-O-Fe superexchange interaction, the lattice elongation along the $b$ axis increases the bond angle of the Fe-O-Fe coupling so that the antiferromagnetic coupling along the $b$ axis is enhanced. As a result, the 4SL spin structure is stabilized.
In Ag$_2$CrO$_2$, one side of the triangle along the $b$ axis is contracted by 0.4\% at $T$=4 K, as shown in Fig. \ref{structure}(c). The antiferromagnetic coupling between NN Cr$^{3+}$ spins is mediated by direct Cr-Cr exchange interaction. Therefore, the contraction along the $b$ axis enhances the antiferromagnetic coupling along the $b$ axis. As shown in Fig. \ref{structure2}, the 5SL structure can be stabilized by enhancing the antiferromagnetic coupling along the $b$ axis. It is very interesting that the triangular lattice is elastic and can expand or contract to stabilize long-range magnetic order. The spin-lattice coupling is a common feature in triangular lattice antiferromagnets in which further-neighbor interactions give rise to multiple-sublattice states.

\section{Origin of the Partially Disordered Phase}
What is the origin of the PD state with 5SL in Ag$_2$CrO$_2$? Further-neighbor interactions should be relevant to such an unusual magnetic state. Mekata {\it et al.} performed Monte Carlo calculations in the triangular lattice Ising magnet with $J_1$, $J_2$, and $J_3$, shown in Fig. \ref{structure}(b).~\cite{mekata93,mekata95} The 5SL state is not stable at $T$=0 K but can appear at finite temperature when $J_2$ and $J_3$ are antiferromagnetic and relatively large, although the region where the phase exists is very narrow. The 4- or 8-sublattice state becomes more stable at low temperatures. 
It was also reported that the Heisenberg model for nearest-neighbor spins including a spin-lattice coupling explains the collinear 4- and 8-sublattice states.~\cite{wang08} As a result of the spin-lattice coupling, the effective spin Hamiltonian becomes an Ising model with nearest-, second-, and third-nearest neighbor coupling, $J(1-c)$, $cJ$, and $cJ$, respectively, where $c$ is a spin-lattice coupling.
In CuFeO$_2$ large $J_2$($\sim$0.52$J_1$) and $J_3$($\sim$0.70$J_1$) were actually observed.~\cite{fishman08} 
This is because the overlap between Fe and O orbitals is relatively large. However, in Ag$_2$CrO$_2$, in which $e_g$ orbitals of the Cr$^{3+}$ ion are not occupied, the overlap of the Cr and O orbitals is small. Therefore, $J_2$ and $J_3$, corresponding to Cr-O-O-Cr super-super exchange interactions, are not expected to be so large compared to those in CuFeO$_2$.
It can be deduced that the non-negligible $J_2$ and $J_3$ originate from the RKKY interaction mediated by the metallic Ag$_2$ layers. The interlayer couplings, which give rise to the long-range magnetic order, is also considered to originate from the RKKY interaction.
An abrupt drop of the resistivity around $T\rm_N$ \cite{yoshida11} indicates a strong coupling between the magnetic CrO$_2$ and metallic Ag$_2$ layers. 
Theoretical studies are desirable to clarify the interesting coupling between the two layers. Further inelastic neutron scattering study is also needed to determine the further-neighbor interactions.

The PD state has been reported in the Ising spin chains, which form the triangular lattice, such as CsCoBr$_3$,~\cite{yelon75} CsCoCl$_3$,~\cite{mekata78}  and RbCoBr$_3$ ~\cite{nishiwaki08} which are antiferromagnetic chain compounds and Ca$_3$CoRhO$_6$  that is a ferromagnetic chain compound.~\cite{niitaka01}  These compounds show PD state with 3 sublattices. In particular, Ca$_3$CoRhO$_6$ shows a PD spin structure, in which the spin arrangement along the $c$ axis is ferromagnetic, as in Ag$_2$CrO$_2$. In Ca$_3$CoRhO$_6$, the PD state appears at $T_1$=90 K and the disordered spins randomly freeze at $T_2$=30 K. The magnetic susceptibility shows a weak ferromagnetic component with a large hysteresis between measurements with field-cool (FC) and zero-field-cool (ZFC) processes below $\sim T_2$. On the other hand, the magnetic susceptibility in Ag$_2$CrO$_2$ shows a weak ferromagnetic component with a tiny hysteresis below $T\rm_N$ down to 2 K,~\cite{yoshida11} indicating that the PD state persists at 2 K. It is probable that the disordered spins fluctuate even at low temperatures due to strong frustration and high two-dimensionality.
As far as we know, this is the first observation of the long-range PD state in quasi-2D triangular lattice antiferromagnet.
The spin-lattice coupling is considered to stabilize the PD state with 5SL, which is expected to appear only at finite temperatures in a limited range of $J_2$ and $J_3$ and has never been observed experimentally.

\begin{figure}
\includegraphics[width=8.0cm]{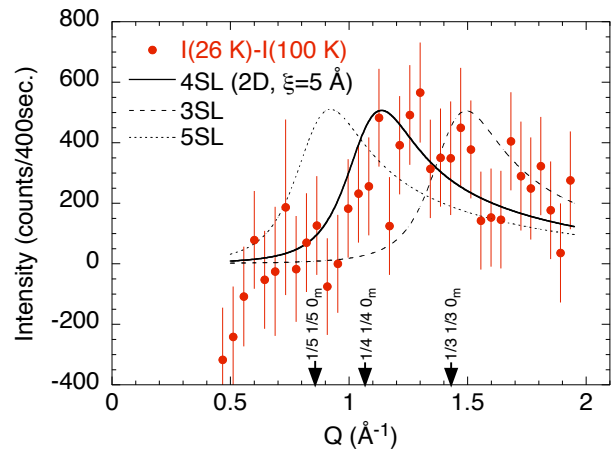}
\caption{(Color online) Magnetic diffuse scattering at $T$=26 K above $T\rm_N$. The scattering at $T$=100 K was subtracted. The negative intensity at low $Q$ probably originates from the paramagnetic scattering that increases at low $Q$ at high temperatures.
The dotted, solid, and broken curves are the results of model calculations for 3-, 4-, and 5-sublattice states with a 2D spin correlation length $\xi$=5 \AA, respectively.}
\label{diffuse}
\end{figure}
We found the interesting spin-lattice coupling that stabilizes the PD state with 5 sublattices. As the next step, it is interesting to clarify whether the 5SL state is preferred above $T\rm_N$. We studied how the spin correlation develops with decreasing temperature.
Figure \ref{diffuse} shows a neutron diffuse scattering spectrum, observed at $T$=26 K, which is 2 K above $T\rm_N$. The spectrum measured at $T$=100 K is subtracted as a background. A broad peak centered around 1.2 \AA$^{-1}$ is observed. The broad peak was simulated using the powder averaged 2D squared Lorentzian. Magnetic form factor was also included in the calculation. Since the peak width is much larger than the instrumental resolution ($\sim$0.04 \AA$^{-1}$), the resolution correction was not performed. The solid curve, which is the result of a model calculation for 4SL state with a 2D spin correlation length $\xi$=5 \AA, reproduces the observed data reasonably well.
Considering the paramagnetic contribution, which is oversubtracted, the agreement would become improved. The model calculations with 3-sublattice (3SL) and 5SL states do not describe the observed data. This result suggests that the spin system prefers 4SL state just above $T\rm_N$. Therefore, the further-neighbor interactions are considered to be already dominant above the structural transition temperature.

In the $J_1$-$J_2$-$J_3$ model, the energies per spin for the 4SL and 5SL phases are evaluated to be $E_4=J_1-J_2+J_3-J_z+J_z'$ and $E_{\rm 5PD}=\frac{1}{5}J_1+\frac{2}{5}J_2+J_3-\frac{4}{5}J_z+\frac{1}{5}J_z'$, respectively, where $J$ is negative for antiferromagnetic interaction. Here, $J_z$ and $J_z'$ represent nearest- and second-nearest-neighbor couplings between the triangular lattice layers, respectively. Our results indicate that $E_4<E_{\rm 5PD}$ above $T\rm_N$. Therefore, the relation of $4J_1-7J_2-J_z+4J_z'<$0 is expected.
The structural transition, which gives rise to an anisotropy in $J_1$, occurs at $T\rm_N$. We discuss the magnetic structure using averaged $J$'s, since there is no theoretical calculation for the $J_1$-$J_2$-$J_3$ model in the distorted lattice.
From the formulas of $E_4$ and $E_{\rm 5PD}$, $J_1$, $J_2$, and $J_z'$ are considered to be influential to select one from 4SL and 5SL states. As mentioned above, $J_1$ enhanced along the $b$ axis stabilizes both 4SL and 5SL states.
In order to stabilize the 5SL state more than the 4SL state, $|$$J_2$/$J_1$$|$ should increase.~\cite{mekata93,mekata95} Since $J_2$ perhaps originates from the RKKY interaction, the abrupt decrease in resistivity around $T\rm_N$ \cite{yoshida11} is considered to change the RKKY interaction and enhance $J_2$. An increase of $J_z'$ also stabilizes the 5SL state.
It is noted that the PD state is stable only at finite temperatures. Other collinear states such as 8-sublattice phase should become stable at $T$=0 K.~\cite{mekata93,mekata95,wang08} Although a PD state with 9-sublattices is predicted between the 4SL and 5SL phases, it was not observed experimentally.

\section{Concluding remarks}
We have studied spin correlations and spin-lattice coupling in a triangular lattice antiferromagnet Ag$_2$CrO$_2$. With decreasing temperature, a short-ranged 4SL spin state first develops. At $T\rm_N$, spin-lattice coupling gives rise to a long-range PD state with 5 sublattices, which is expected to appear only in limited ranges of $J_2$, $J_3$, and temperature.
Since this is the first observation of the long-range PD phase in quasi-2D triangular lattice magnet, theoretical progress in this field is highly desirable.

\section*{Acknowledgments}
We would like to thank Profs. Y. Maeno and Y. Motome for stimulating discussions. The work at ORNL was sponsored by the Scientific User Facilities Division and Materials Sciences and Engineering Division, Office of Basic Energy Sciences, U. S. Department of Energy.

\end{document}